# From a set of parts to an indivisible whole.
# Part II: Operations in an open comparative mode

Leonid Andreev

Equicom, Inc., 10273 E Emily Dr, Tucson, AZ 85730, U.S.A.
E-mail: equicom@matrixreasoning.com

May 5, 2008

## Abstract

This paper describes a new method, HGV2C, for analysis of patterns, which represents a new modification of the previously described method for non-probabilistic hypothesis generation and verification (HGV). The HGV2C method involves the construction of a 'computer ego' (CE) based on an individual object that can be either a part of the system under analysis or a newly created object based on a certain hypothesis (model). The CE provides the capability to analyze data from a specific standpoint, e.g. a specific object's "viewpoint". This new approach to knowledge representation is demonstrated on the example of population pyramids of 220 countries.

The CE is constructed from two identical copies (clones) of a query object, and its functioning mechanism involves two elements: a hypothesis-parameter (HP) and infothyristor (IT). HP is a parameter that is introduced into an existing set of parameters describing the objects of a given system. In the simplest version of the HGV2C method described in this paper, the HP value for one of the clones of a query object is set to equal 1, whereas for another clone it is more than 1 (the difference between the two values is denoted as $\Delta$). The IT is based on the previously described algorithm of iterative averaging which provides data processing in such a way that a dataset under processing undergoes a division into two alternative groups without outliers. The division results are logical and in accordance with a natural hierarchy inherent in any functioning system. The IT performs three major functions: 1) computation of a similarity matrix for the group of three objects including two clones of the query object and the target object; 2) division of the group of three objects into two alternative subgroups; and 3) a successive increase of the HP weight in the totality of all the parameters, i.e. HP multiplication. Initially, both clones of the query object appear together in one of the subgroups as all of their parameter values, except the HP, are identical. At a certain point of the HP multiplication, one of the clones moves to group of the target object. A respective number of the HP multiplications that results in such regrouping represents the dissimilarity ($D$) between the query and target objects. Sensitivity of determination of $D$ is extremely high, and the product of $D$ multiplied by $\Delta$ is strictly constant and linearly increases as the $\Delta$ value decreases. The $D$ value represents the sum of increments that correspond to each individual parameter.

The HGV2C analysis of the age and sex pyramids of 220 countries showed that any population pyramid represents the additive sum of two components – the exponential and uniform ones. In an exponential type pyramid, the share of each age cohort decreases, at a constant rate, relative the immediately preceding cohort (the most characteristic example is the population pyramid of Uganda); whereas in a uniform type pyramid, the shares of all age cohorts are constant (the most characteristic example is Monaco). The validity of this conclusion has been demonstrated on the totality of population pyramids of practically all of the countries of the world. This discovery has led to development of an index ($MU$) that describes the entire diversity of the population pyramids based on a quantitative criterion that correlates with the countries' welfare and can be used as an efficient analytical tool in demographic studies.

# 1. Introduction

This is a second article of the series of papers, including the recently published one [1], on various forms of algorithmic implementation of the central idea of holism - the possibility of synthesis of an indivisible whole from a set of its scattered elements. Practical realization of this idea has become achievable due to our discovery of a new phenomenon - any data system subjected to iterative averaging undergoes a division that produces two alternative subgroups of elements, without outliers [2]. Previously [1], we provided a detailed description of an algorithm for iterative averaging (the algorithm of evolutionary transformation of similarity matrices, ETSM) which, along with specially designed metrics [2] and a new method for computation of similarity matrices [3] and techniques for construction of hierarchical trees and dendrograms [2] provides an efficient instrument for routine analysis of systems of any kind, during which the synthesis of an indivisible whole occurs as an antecedent of analysis of the functional role of its individual elements. This quite an unusual approach to discovery of the nature of objects and phenomenon and to knowledge representation, called by us 'matrix reasoning' [2], involves a subdivision of each newly formed subgroup through iterative averaging, which ultimately results in hierarchical structures that reflect the relationships between the elements in a system under study.

Previously [1], we pointed out two conditions that are obligatory in order to discover the natural hierarchy in a system under analysis: (a) the elements of a system need to be allowed to interact with each other based on the principle of self-organization; and (b) a system under study has to be a closed system. We showed that self-organization can be achieved through a spontaneous overall cross-averaging of a system's elements, which is provided by the ETSM-algorithm. In the course of the ETSM-processing, the very first of the iterative transformations turns any system under processing into a closed type system that does not allow an addition of new elements or removal of any of its existing elements as such changes would inevitably result in a drastic transformation of the original state of the input data system, i.e. the latter would cease to exist as a wholesome object originally taken for the investigation.

The methodology based on iterative averaging (IA) creates new capabilities for development of a whole series of novel approaches to synthesis of an indivisible whole from a set of its elements. In this paper, we present one of such approaches, previously disclosed in [4]. It is also based on the use of the algorithm for iterative transformation of similarity matrices; however, it has its own specifics and offers significant advantages in solution of a variety of practical and research problems in computer science. The main distinction of this approach is a capability to conduct open mode comparative analysis, i.e. without restrictions on the addition of new objects to an input database or removal of any of the existing objects from an input database. In essence, what is described below is a universal method for pattern recognition.

# 2. Algorithmic principles of the HGV2C method

The HGV2C method is a new modification of the previously disclosed HGV method for non-probabilistic hypothesis generation and verification [4]. It is easy for implementation and interpretation. As well as the HGV method, the HGV2C utilizes a number of fundamentally new techniques and elements - a 'computer ego', 'hypothesis-parameter', and 'infothyristor'. The latter two provide for the functioning of the 'computer ego' component. The necessity in creating a capability to enable a computer program have its own point of view on information under processing has arisen since the very beginning of computer science and is directly connected with the problem of artificial intelligence. Unlike the IA-method [1, 2] where interrelations between the elements of a whole are determined by an initial set of elements and established solely by the IA-algorithm with no involvement of the operator's will, in conventional data-processing, a strategy is pre-set in a respective computer program, i.e. in a certain sequence of elementary operations used for evaluation and comparison of individual data points. In the meantime, as well as it is the case with visual perception, quality and results of a data-processing analysis depend not only on the technical aspect of the process, e.g. visual acuity as in the analogy with visual perception, but also on a task-specific response of the observer, e.g. the expectation of a certain result, the angle of vision, the *a priori* set weights of individual details of the complex picture being assessed, etc. In the context of computer-based information processing, it should be appropriate to call such a response a computer ego (CE), i.e. a capability, given to the computer through certain means, to provide a specific and complex response to a dataset under analysis. Examples of CE are query-specific search engines, back-propagated artificial neural networks that are trained by humans and thus acquire a CE to perform specific tasks, etc. Most of such methods provide merely a detector for identification of certain unique properties of objects under study, whereas the capability to not only detect but also assess the detected properties within a discrete rate scale is by far more valuable and productive.

The mechanism of the HGV2C method is as follows. A computer ego is created based on one of the objects of a database under analysis or an object created based on a certain hypothesis (model). It serves as a query object (query) and is represented in the computer ego in two identical copies (clones) – α and β. In addition to the existing parameters of the objects under analysis, we introduce an additional parameter, referred to as a hypothesis-parameter (HP). In the simplest version of the HGV2C method described in this paper, the HP value for α-clone was set to equal 1, whereas for β-clone it was greater than 1. The difference between HP values for α- and β-clones is denoted as Δ. The HP values for all other objects of the system under analysis are set to be 1.

The engine of the method is the so-called infothyristor that performs three functions: 1) computes a similarity matrix for α- and β-clones of a query object $Q$ and a target object $T$; 2) provides the iterative averaging processing of the similarity matrix which results in formation of two alternative subgroups; and 3) provides multiplication of HP, i.e. successively increases its weight (number of copies) in the totality of all the parameters describing the objects under analysis.

Assume that we want to determine the dissimilarity coefficient ($D$) between $Q$ and a certain target object, $T$, of the database. If we subject the three objects – two clones of $Q$ ($Q_\alpha$ and $Q_\beta$) and $T$ to processing by ETSM-algorithm, we will obtain a two-branch tree where both clones of $Q$ will be on one of the branches, and $T$ will be on another branch. This will happen because all of the parameters, except for the hypothesis-parameter (HP) values, of the $Q$ clones are identical and in some way or another differ from the object $T$ parameters. If we start multiplication of the HP, i.e. increasing its weight in the totality of the parameters, then, after a certain number of multiplications, clone $Q_\alpha$ will move to the branch of object $T$. The number of multiplications that causes $Q_\alpha$ move to branch $T$ represents the dissimilarity $D$ of $Q$ to $T$. If a target object is identical to $Q$, then $D$ will equal 1. If the HP values set for $Q_\alpha$ and $Q_\beta$ are too close to each other (the Δ value is too low), it will be much harder to cause clone $Q_\alpha$ move to branch $T$, which means that the $D$ values will be much higher, and therefore the sensitivity of computation of $D$ will be greater. Sensitivity of computation of $D$ by using the infothrystor is determined by the equation:

$$D = K_{QT} \cdot \Delta^{-1} \qquad (1),$$

where $K_{QT}$ is a constant for a given pair of $T$ and $Q$, and Δ is the difference between HP values of $Q$'s two clones, $Q_\alpha$ and $Q_\beta$. That is, constant $K_{QT}$ is a dissimilarity coefficient at Δ = 1. Thus, the product of multiplication of $D$ by Δ is constant and equals a certain coefficient that depends on a given set of parameters of $T$ and $Q$. As it will be shown further in this paper, despite the known opposite regularity, the higher the analysis sensitivity, the greater is the accuracy of $D$ computation.

The $D$ value is strictly additive – it equals the sum of the D increments produced by each of the individual parameters of object $T$. This property of $D$ does not depend on which of the two metrics – R or XR [1, 2] – is used for computation of the similarity matrix. Thus, the $D$ values computed for $T$ in relation to $Q$ based on individual parameters or different groups of parameters can be stored in a special database and can be combined and used as necessary. The processing time for each of the objects of a system under analysis is constant and depends on a computer's performance characteristics.

Pattern recognition or pattern classification is one of the most demanded solutions in the present-day computer science [5, 6]. The application areas for pattern recognition techniques are so numerous and vast that it would be virtually impossible to enumerate all of them (such as, for example, computer vision, speech, character, text, etc. recognitions; medical image analysis, biometrics, trading pattern analysis, etc.); and it would be equally hard to name at least one field of modern knowledge and practice in which the utilization of efficient pattern recognition technology would not promise significant success.

Modern pattern recognition techniques are based on the use of neural networks, hidden Markov models, Bayesian networks, artificial intelligence (expert systems and machine learning), cluster analysis, mathematical statistics (hypothesis testing and parameter estimation, discriminant analysis and feature extraction), etc. This powerful set of data processing tools is used in pattern recognition methodology not only with the purpose of recognition of patterns in data but also as the means for understanding of how patterns are formed, what external or internal factors are responsible for formation of given patterns, and what patterns may form in unforeseen situations. The capability of pattern formation forecasting by interpolation and extrapolation – which constitutes intelligent pattern recognition – is quite a rare feature in the heretofore known methods for pattern recognition.

The most widely used technology for pattern recognition is the so-called fingerprinting [7] or "hieroglyphic method", when a target pattern is compared to a pattern whose physical characteristics are *a priori* known. This type of pattern recognition is based on a search for an image that is similar to a target pattern represented in the form of a certain unique symbol. However effective such an approach may look in practice, it has considerable limitations in the ability to predict the behavior of patterns depending on various factors. For instance, the shape of population pyramids can significantly change over a period of time, and a fingerprint comparison between such pyramids that have undergone changes in time does not always allow an intelligent analysis of the causes of such changes. The same is true for most of the other known techniques for pattern recognition. Therefore, the need in truly intelligent methods for pattern recognition is continually growing. There is a need in methods that allow thorough investigations into mechanisms of pattern formation under the effect of natural factors.

Below we will demonstrate that the HGV method, in its HGV2C implementation presented in this work, represents an intelligent method for pattern recognition which allows analysis of patterns of any nature and origin, because it provides the response additivity, an ideal fusibility of objects' attributes, and a high scalability.

## 3. Comparative analysis of population pyramids

To demonstrate the intelligence of HyGV2C method, we have chosen population pyramids as objects of analysis, for the following reasons. 1. Demographic data are publicly available from reliable sources (e.g. U.S. Census Bureau) and are regularly updated. 2. Population pyramids change over a course of time, and each country population pattern has its own dynamics of changing, which allows investigations into dynamics into age and sex distributions in population pyramids under the influence of various natural factors. 3. Population pyramids usually contain 34 parameters in the form of five-year cohorts (17 parameters for each of the two sex groups of country populations), which makes their patterns sufficiently complex for the purpose of demonstration of the application potentials of the proposed method for pattern recognition. 4. Pyramid shapes significantly differ for different countries (see, e.g., Fig. 1) due to the impact of a whole range of factors, such as national specifics which is closely connected with geographical location, predominant religion (see, e.g. [1]), specifics of economic and political development, influence of neighboring countries, etc. 5. Population pyramids reflect highly complex processes of cooperative interrelations between the whole world populations which is especially relevant in the situation of globalization. 6. Population pyramids have been in the focus of detailed investigations by the demographic research community since long ago, and validity of any new findings can be easily confirmed or rejected based on prior knowledge and well-established views in this field. Different age groups have different dynamics of birth and death rates, population migration, etc., and still all these processes are cooperative and closely interrelated. Therefore, population pyramids represent a perfect subject for evaluation of the intelligence of any given method for pattern recognition. Fig. 1 above shows population pyramids of 16 countries demonstrating widely varying patterns.

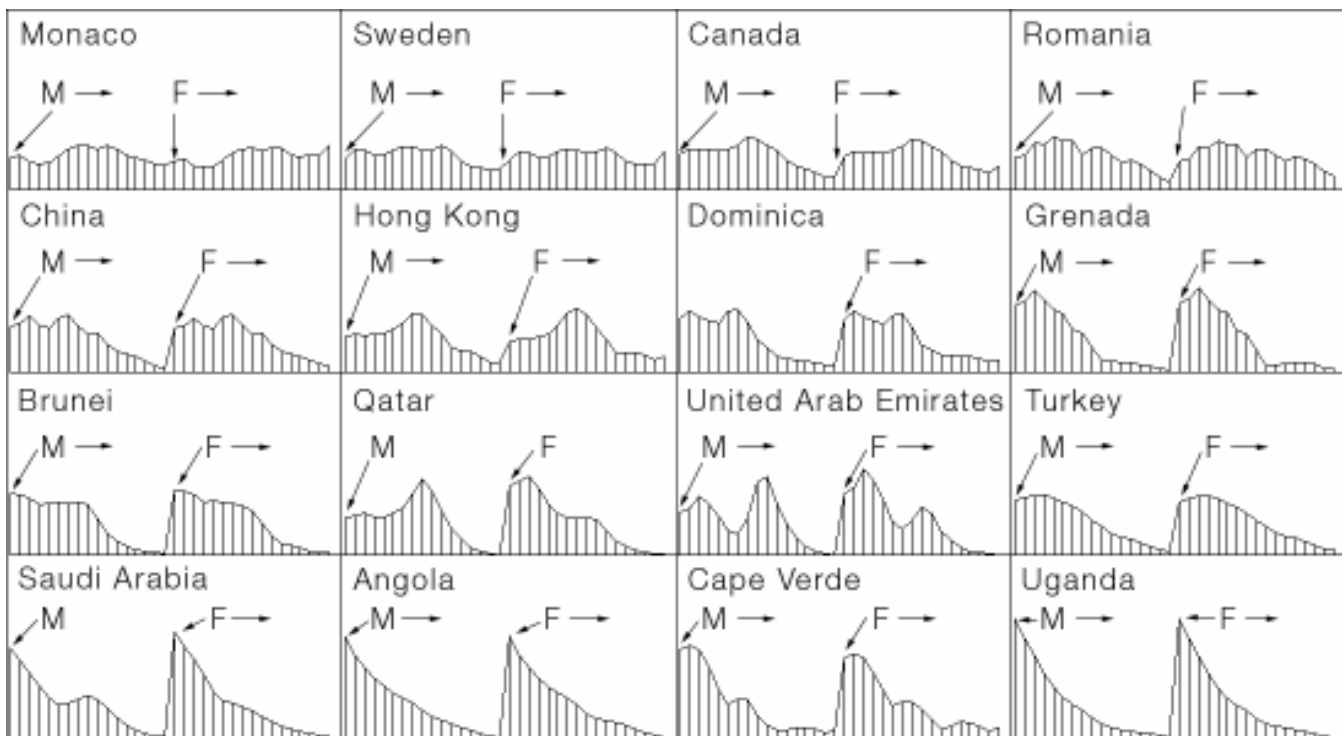

FIG. 1. Population pyramids of selected countries. X-axis shows 5-year age group intervals; point M is the start of the male population section of a pyramid; point F, female population. Y-axis shows percentage of each age group in a given population.(Data source: U.S. Census Bureau, International Data Base, IDB Summary Demographic Data, 2000. http://www.census.gov/ipc/www/idbsum.html).

### 3. 1. Objects and conditions of analysis

We used demographic data on population age and sex in 220 countries (year 2000 data by U.S. Census Bureau, International Data Base, IDB Summary Demographic Data, 2000. http://www.census.gov/ipc/www/idbsum.html)[1]. The input data table contained data on 17 age groups of each sex: 16 of the 5-year interval groups (00 – 04, 05 – 09, and so on) and the 80+ years of age groups. Further, the data were converted into relative percentage of male and female of each age group relative to a country's total population. All the data processing was done with the use of software *MeaningFinder 2.3* developed by Equicom, Inc. The analysis was conducted as follows. To 34 parameters describing each of the population pyramids under analysis, we added one more – hypothetical - parameter, HP. We selected a query object, i.e. a population pyramid to which all other population pyramids were compared and based on which the CE was created by using two duplicates (clones) of the query object. In the simplified variant of analysis demonstrated in this work, the HP value for one of the clones was set to 1, and for the other, greater than 1. The HP values assigned to all other objects under analysis were set to 1. Engagement of the infothyristor automatically starts the computation of a similarity matrix for the two clones of the query object and the target object. In this study, we used the R-metric [1, 2] which represents a ratio between the lower and the higher values of each parameter. The similarity matrix for the three objects was computed by the method of hybridization of monomer similarity matrices [1, 3] and was then processed by the ETSM algorithm [1, 2]. The very first transformation of the similarity matrix produces a two-branch tree showing both clones of the query object on one of the branches and the target object on the other. This happens because the clones are practically identical and differ from each other only by the values of the hypothesis-parameter. After each iterative averaging cycle, the weight of the hypothesis-parameter in relation to the other parameters is increased by a certain number of multiplications. When the hypothesis-parameter weight reaches a certain level, the structure of the hierarchical tree changes: the query-object's clone whose hypothesis-parameter value equals 1 moves to the branch that previously had only the target object on it. The number of the HP multiplications which results in the clone's movement to the target object branch of the hierarchical tree is recorded as a dissimilarity coefficient $D$ between a query and target objects.

---

[1] Here, the term 'countries' applies to both sovereign states and relatively autonomous populations, such as, for instance, Isle of Man, the Island of Jersey, Faeroe Islands, Gaza Strip, etc.

## 3.2. Sensitivity of determination of dissimilarity coefficients

Sensitivity of determination of dissimilarity coefficient $D$ is described by Eq. (1) and depends on coefficient $K_{QT}$. Essentially, $K_{QT}$ represents a certain holistic characteristic of a target pattern evaluated in comparison to a query pattern. Table 1 shows the $K_{QT}$ values for population pyramids of some of the countries which were computed upon the analysis where query objects were population pyramids of Monaco, Argentina, and Uganda. The $K_{QT}$ values were computed at different $\Delta$ values that represent the difference between HP values of two clones. As is seen from the data in Table 1, the $K_{QT}$ values are the more stable as the absolute values of $\Delta$ are lower, hence the determination of $D$ is more sensitive. The product of multiplication of $D$ by the $\Delta$ values, i.e. constant $K_{QT}$ is practically stable despite that the $\Delta$ values vary up to 5 orders of magnitude, which is the more so remarkable as the $K_{QT}$ values for different population pyramids vary by more than 7 times.

**Table 1.** $K_{QT}$ values computed for population pyramids of various countries by using population pyramids of Monaco, Argentina and Uganda as query objects. Data source: U.S. Census Bureau, International Data Base, IDB Summary Demographic Data, 2000. http://www.census.gov/ipc/www/idbsum.html).

| $\Delta$ values | $K_{QT}$ values (see Eq. (1). Query object - Monaco | | | | | | | | |
|---|---|---|---|---|---|---|---|---|---|
| | Sweden | Japan | Austria | Russia | Argentina | China | Afghanistan | Nigeria | Uganda |
| $10^{-1}$ | 6.230 | 6.860 | 8.180 | 13.500 | 17.960 | 22.260 | 39.330 | 40.230 | 47.540 |
| $10^{-2}$ | 5.968 | 6.567 | 7.829 | 12.931 | 17.202 | 21.316 | 37.664 | 38.529 | 45.537 |
| $10^{-3}$ | 5.941 | 6.537 | 7.794 | 12.873 | 17.125 | 21.220 | 37.496 | 38.356 | 45.333 |
| $10^{-4}$ | 5.938 | 6.534 | 7.790 | 12.867 | 17.117 | 21.210 | 37.479 | 38.339 | 45.313 |
| $10^{-5}$ | 5.938 | 6.533 | 7.790 | 12.867 | 17.116 | 21.209 | 37.477 | 38.337 | 45.311 |
| $10^{-6}$ | 5.938 | 6.533 | 7.790 | 12.867 | 17.116 | 21.209 | 37.477 | 38.377 | 45.310 |
| $10^{-7}$ | 5.938 | 6.533 | 7.790 | 12.866 | 17.116 | 21.209 | 37.477 | 38.337 | 45.310 |

| $\Delta$ values | $K_{QT}$ values (see Eq. (1). Query object - Argentina | | | | | | | | |
|---|---|---|---|---|---|---|---|---|---|
| | Sweden | Japan | Austria | Russia | Argentina | China | Afghanistan | Nigeria | Uganda |
| $10^{-1}$ | 12.040 | 13.000 | 11.850 | 9.620 | 8.060 | 21.400 | 22.430 | 17.960 | 30.250 |
| $10^{-2}$ | 11.527 | 12.447 | 11.347 | 9.211 | 7.718 | 20.507 | 21.478 | 17.202 | 28.971 |
| $10^{-3}$ | 11.475 | 12.391 | 11.296 | 9.169 | 7.682 | 20.415 | 21.381 | 17.125 | 28.841 |
| $10^{-4}$ | 11.469 | 12.385 | 11.291 | 9.165 | 7.678 | 20.406 | 21.372 | 17.117 | 28.828 |
| $10^{-5}$ | 11.469 | 12.385 | 11.290 | 9.165 | 7.678 | 20.405 | 21.371 | 17.116 | 28.826 |
| $10^{-6}$ | 11.469 | 12.385 | 11.290 | 9.165 | 7.678 | 20.405 | 21.371 | 17.116 | 28.826 |
| $10^{-7}$ | 11.469 | 12.385 | 11.290 | 9.165 | 7.678 | 20.405 | 21.371 | 17.116 | 28.826 |

| Δ values | $K_{QT}$ values (indexes) (see Eq. (1). Query object - Uganda | | | | | | | | |
|---|---|---|---|---|---|---|---|---|---|
|  | Sweden | Japan | Austria | Russia | Argentina | China | Afghanistan | Nigeria | Uganda |
| $10^{-1}$ | 41.700 | 43.050 | 41.670 | 36.670 | 30.250 | 28.770 | 9.160 | 8.110 | 47.540 |
| $10^{-2}$ | 39.940 | 41.236 | 39.911 | 35.117 | 28.971 | 27.552 | 8.769 | 7.762 | 45.537 |
| $10^{-3}$ | 39.761 | 41.051 | 39.732 | 34.960 | 28.841 | 27.429 | 8.730 | 7.762 | 45.333 |
| $10^{-4}$ | 39.743 | 41.033 | 39.714 | 34.944 | 28.828 | 27.416 | 8.726 | 7.723 | 45.313 |
| $10^{-5}$ | 39.741 | 41.031 | 39.713 | 34.943 | 28.826 | 27.415 | 8.725 | 7.723 | 45.311 |
| $10^{-6}$ | 39.741 | 41.031 | 39.713 | 34.942 | 28.826 | 27.415 | 8.725 | 7.723 | 45.310 |
| $10^{-7}$ | 39.741 | 41.031 | 39.713 | 34.942 | 28.826 | 27.415 | 8.724 | 7.723 | 45.310 |

### 3.3. Specificity and additivity of $K_{QT}$ coefficients

As is seen from Table 1, the $K_{QT}$ coefficient is a specific characteristic of a given pattern and depends on a set of parameters describing the pattern (in this example, the set of parameters was percentages of different age cohorts in a population pyramid of a given country). The $K_{QT}$ coefficient value significantly varies depending on which population pyramid is used as a query pattern.

Thus established sets of $K_{QT}$ coefficients reflect a response of the "computer ego" based on the clones of a query objects. In other words, a signal generated upon processing of target patterns is based on the "viewpoint" of the computer ego. The $K_{QT}$ values depend on how similar a target pattern is to a query pattern that underlies the computer ego. The more similar the target and query patterns, the lower the $K_{QT}$ values.

Another unique peculiarity of the $K_{QT}$ coefficient is its being ultimately additive. It equals the sum of partial values of $K_{QT}$ produced by each individual parameter. In case of population pyramids, the $K_{QT}$ coefficient computed for an entire given pattern equals the sum of $K_{QT}$ values computed for each of the 34 cohorts. To illustrate the above, Table 2 shows $K_{QT}$ coefficients for population pyramids of 155 countries, including $K_{QT}$ values computed individually for female and male populations of each country, the sum of those values, and the $K_{QT}$ values determined for entire population pyramids (columns 1 through 4). In this analysis, the query object was Monaco.

As is seen from Table 2, the values in columns 3 and 4 ideally coincide. The maximum difference of 0.001 is due only to the rounding of the values. The ratio of $K$ for the female component of a pyramid to K for the mail component of the pyramid varies for different countries within quite a wide range. For instance, the ratio for Saudi Arabia, Kuwait and Croatia is 0.77 – 0.79; for Israel, Argentina, Egypt, and Switzerland 1.00 – 1.01; whereas for Latvia it is 1.90; which shows a variation by 2.5 times. Nonetheless, on the example of 220 population pyramids, one can see a good correlation between these two values of the $K_{QT}$ index. Fig. 2 shows a linear dependence between the values of $K_{UT}(f)$ for the female population and $K_{UT}(m)$ for the male population of each pyramid. The results were obtained by using Uganda as a query object. The correlation shown in Fig. 2 is well described by a simple equation:

$$K_{UT}(f) = 0.30 + 1.025 K_{UT}(m) \qquad (2)$$

The data shown in Fig. 2 indicate that among 220 countries, Uganda and Monaco have the highest level of dissimilarity to each other. As is seen in Table 2, the sums of the coefficients $K_{QT}$ computed upon using these two countries as query objects is close to a constant value whose average is 47.2 with an average standard deviation of 0.94. The highest deviations from the average value are observed for oil-producing countries Kuwait, Qatar and United Arab Emirates, where the population dynamics is regulated by the government. The sum of the $K_{QT}$ values is constant only when query objects are Uganda and Monaco, the two countries that, in the set of 220 population pyramids, represent polar opposites. For instance, upon use of Monaco and Argentina as query objects (Table 2, column 7), the sum of the $K_{QT}$ coefficients for 155 countries varies by more than 4 times. Fig. 3 shows a linear

Table 2. $K_{QT}$ values computed for population pyramids of 155 countries. Data source: U.S. Census Bureau, International Data Base, IDB Summary Demographic Data, 2000. http://www.census.gov/ipc/www/idbsum.html).

| Country | $K_{QT}$ values | | | | | | | |
|---|---|---|---|---|---|---|---|---|
| | 1 | 2 | 3 | 4 | 5 | 6 | 7 | 8 |
| Afghanistan | 17.624 | 19.853 | 37.477 | 37.477 | 8.725 | 46.2 | 57.9 | 17.9 |
| Albania | 11.353 | 11.524 | 22.877 | 22.878 | 23.541 | 46.4 | 29.3 | 50.7 |
| Algeria | 16.272 | 17.088 | 33.360 | 33.359 | 14.628 | 48.0 | 49.7 | 30.0 |
| Angola | 18.800 | 19.736 | 38.536 | 38.537 | 7.857 | 46.3 | 60.0 | 16.0 |
| Argentina | 8.558 | 8.559 | 17.117 | 17.116 | 28.828 | 45.9 | n/a | 63.1 |
| Australia | 6.073 | 5.340 | 11.413 | 11.413 | 35.322 | 46.7 | 27.6 | 76.3 |
| Austria | 4.196 | 3.594 | 7.790 | 7.790 | 39.713 | 47.5 | 19.1 | 84.5 |
| Azerbaijan | 13.031 | 11.892 | 24.923 | 24.924 | 22.488 | 47.4 | 34.5 | 47.4 |
| Bahrain | 15.240 | 17.446 | 32.686 | 32.687 | 17.080 | 49.8 | 51.4 | 33.9 |
| Bangladesh | 15.782 | 18.382 | 34.164 | 34.164 | 13.507 | 47.7 | 51.2 | 27.7 |
| Belarus | 6.920 | 4.841 | 11.761 | 11.762 | 35.313 | 47.1 | 19.8 | 75.8 |
| Belgium | 3.415 | 3.348 | 6.763 | 6.763 | 39.544 | 46.3 | 18.0 | 86.4 |
| Benin | 21.751 | 20.514 | 42.265 | 42.265 | 4.851 | 47.1 | 67.5 | 9.1 |
| Bermuda | 5.867 | 6.363 | 12.230 | 12.230 | 36.426 | 48.7 | 21.1 | 75.6 |
| Bhutan | 15.364 | 16.886 | 32.250 | 32.250 | 13.711 | 46.0 | 47.6 | 29.2 |
| Bolivia | 15.660 | 15.869 | 31.529 | 31.528 | 14.775 | 46.3 | 46.0 | 31.4 |
| Botswana | 16.740 | 17.584 | 34.324 | 34.324 | 13.993 | 48.3 | 53.5 | 24.7 |
| Brazil | 13.371 | 12.795 | 26.166 | 26.166 | 21.722 | 47.9 | 35.8 | 45.2 |
| Brunei | 15.901 | 17.873 | 33.774 | 33.774 | 15.167 | 48.9 | 52.1 | 30.4 |
| Burkina Faso | 20.882 | 19.475 | 40.357 | 40.356 | 6.432 | 46.8 | 63.8 | 12.7 |
| Burma | 14.327 | 14.676 | 29.003 | 29.002 | 19.440 | 48.4 | 41.4 | 39.9 |
| Burundi | 19.995 | 19.276 | 39.271 | 39.270 | 6.402 | 45.7 | 62.2 | 13.0 |
| Cambodia | 18.393 | 16.968 | 35.361 | 35.361 | 10.681 | 46.0 | 54.4 | 22.5 |
| Canada | 5.270 | 5.711 | 10.981 | 10.981 | 36.545 | 47.5 | 19.2 | 77.7 |
| CAR | 17.834 | 17.766 | 35.600 | 35.600 | 10.547 | 46.2 | 54.2 | 22.1 |
| Chad | 20.372 | 19.165 | 39.537 | 39.537 | 6.147 | 45.7 | 62.4 | 12.3 |
| Chile | 10.272 | 10.086 | 20.358 | 20.358 | 26.291 | 46.7 | 25.4 | 56.6 |
| China | 10.187 | 11.023 | 21.210 | 21.209 | 27.415 | 48.6 | 28.9 | 56.6 |
| Columbia | 13.963 | 14.277 | 28.240 | 28.239 | 19.313 | 47.6 | 40.3 | 42.1 |
| Congo (B) | 19.026 | 18.265 | 37.291 | 37.291 | 9.305 | 46.6 | 57.5 | 19.1 |
| Congo (K) | 20.929 | 20.448 | 41.377 | 41.377 | 4.566 | 45.9 | 66.0 | 8.7 |
| Costa Rica | 13.053 | 13.778 | 26.831 | 26.831 | 20.185 | 47.0 | 37.3 | 42.7 |
| Croatia | 3.877 | 4.889 | 8.776 | 8.775 | 37.498 | 46.3 | 18.5 | 81.9 |
| Cuba | 7.506 | 8.293 | 15.799 | 15.799 | 33.557 | 49.4 | 22.2 | 68.5 |
| Cyprus | 6.620 | 6.934 | 13.554 | 13.554 | 32.334 | 45.9 | 17.8 | 71.7 |
| Czech Rep. | 5.333 | 4.535 | 9.868 | 9.868 | 37.574 | 47.4 | 19.0 | 80.0 |
| Denmark | 3.489 | 3.740 | 7.249 | 7.248 | 39.096 | 46.3 | 17.9 | 85.4 |
| Djibouti | 17.046 | 19.570 | 36.616 | 36.616 | 9.551 | 46.2 | 56.4 | 19.8 |
| Ecuador | 14.938 | 15.834 | 30 772 | 30.773 | 16.342 | 47.1 | 44.5 | 34.2 |
| Egypt | 15.957 | 15.798 | 31.755 | 31.756 | 15.330 | 47.1 | 46.5 | 32.1 |
| El Salvador | 15.280 | 15.372 | 30.652 | 30.652 | 16.133 | 46.8 | 44.2 | 34.0 |
| Eritrea | 17.858 | 18.825 | 36.683 | 36.683 | 10.412 | 47.1 | 56.3 | 21.3 |
| Estonia | 6.021 | 3.943 | 9.964 | 9.964 | 36.792 | 46.8 | 18.8 | 79.5 |
| Ethiopia | 19.281 | 19.936 | 39.219 | 39.217 | 6.331 | 45.6 | 61.7 | 12.9 |
| Finland | 4.142 | 3.539 | 7.681 | 7.681 | 38.737 | 46.4 | 18.2 | 84.5 |
| France | 4.101 | 4.033 | 8.134 | 8.134 | 37.961 | 46.1 | 17.6 | 83.3 |

| Country | | | | | | | | |
|---|---|---|---|---|---|---|---|---|
| Gabon | 11.959 | 13.144 | 25.103 | 25.103 | 20.783 | 45.9 | 34.8 | 45.1 |
| Gambia, The | 18.419 | 20.011 | 38.430 | 38.429 | 7.267 | 45.7 | 60.1 | 14.8 |
| Georgia | 7.123 | 6.013 | 13.136 | 13.136 | 33.486 | 46.6 | 20.0 | 72.5 |
| Germany | 3.949 | 3.087 | 7.036 | 7.036 | 41.316 | 48.4 | 20.1 | 86.5 |
| Ghana | 17.589 | 18.308 | 35.897 | 35.897 | 10.541 | 46.4 | 54.7 | 22.0 |
| Gibraltar | 4.166 | 4.575 | 8.741 | 8.741 | 37.407 | 46.2 | 18.1 | 81.9 |
| Greece | 3.427 | 3.750 | 7.177 | 7.177 | 40.117 | 47.3 | 18.6 | 85.8 |
| Greenland | 10.951 | 12.164 | 23.115 | 23.114 | 25.942 | 49.1 | 35.6 | 53.0 |
| Guinea | 19.563 | 18.84 | 38.403 | 38.403 | 7.902 | 46.3 | 59.8 | 16.2 |
| Guinea-Bissau | 18.729 | 18.512 | 37.241 | 37.241 | 9.217 | 46.6 | 57.4 | 18.9 |
| Haiti | 16.810 | 19.165 | 35.975 | 35.975 | 11.504 | 47.5 | 55.1 | 23.4 |
| Honduras | 17.778 | 18.220 | 35.998 | 35.998 | 10.55 | 46.6 | 55.0 | 21.9 |
| Hong Kong | 6.404 | 8.028 | 14.432 | 14.432 | 35.818 | 50.3 | 22.9 | 71.8 |
| Hungary | 5.225 | 4.015 | 9.240 | 9.239 | 37.714 | 47.0 | 18.5 | 81.2 |
| Iceland | 6.244 | 7.205 | 13.449 | 13.449 | 32.708 | 46.2 | 17.8 | 71.4 |
| India | 13.365 | 15.009 | 28.374 | 28.374 | 15.868 | 44.2 | 39.8 | 41.0 |
| Indonesia | 14.204 | 14.556 | 28.760 | 28.760 | 19.255 | 48.0 | 41.1 | 39.8 |
| Iran | 14.930 | 16.906 | 31.836 | 31.837 | 16.021 | 47.9 | 46.6 | 33.0 |
| Iraq | 19.078 | 19.854 | 38.932 | 38.933 | 8.559 | 47.5 | 60.8 | 17.1 |
| Ireland | 6.699 | 7.157 | 13.856 | 13.856 | 32.432 | 46.3 | 17.6 | 70.7 |
| Israel | 9.022 | 9.047 | 18.069 | 18.070 | 28.155 | 46.2 | 39.2 | 61.3 |
| Italy | 3.040 | 2.941 | 5.981 | 5.980 | 42.372 | 48.4 | 19.6 | 88.7 |
| Japan | 3.348 | 3.186 | 6.534 | 6.533 | 41.031 | 47.6 | 18.9 | 87.3 |
| Jordan | 17.648 | 19.023 | 36.671 | 36.671 | 11.713 | 48.4 | 56.3 | 23.4 |
| Kazakhstan | 12.095 | 9.860 | 21.955 | 21.954 | 24.573 | 46.5 | 29.3 | 52.9 |
| Kenya | 20.315 | 20.304 | 40.619 | 40.619 | 7.090 | 47.7 | 64.2 | 13.9 |
| Kyrgyzstan | 14.981 | 13.365 | 28.346 | 28.350 | 17.959 | 46.3 | 39.8 | 38.5 |
| Korea North | 12.687 | 9.979 | 22.666 | 22.666 | 25.86 | 48.5 | 32.1 | 53.5 |
| Korea South | 10.533 | 9.641 | 20.174 | 20.175 | 28.954 | 49.1 | 28.3 | 59.1 |
| Kuwait | 15.768 | 20.385 | 36.153 | 36.154 | 15.045 | 51.2 | 57.3 | 28.9 |
| Laos | 17.751 | 18.187 | 35.938 | 35.938 | 10.156 | 46.1 | 54.9 | 21.2 |
| Latvia | 5.850 | 3.082 | 8.932 | 9.652 | 37.623 | 47.3 | 19.3 | 80.4 |
| Liberia | 16.728 | 17.981 | 34.709 | 34.709 | 11.173 | 45.9 | 52.6 | 23.6 |
| Lichtenstein | 5.558 | 5.922 | 11.480 | 11.480 | 36.922 | 48.4 | 20.3 | 77.0 |
| Libya | 16.526 | 18.533 | 35.059 | 35.058 | 13.645 | 48.1 | 53.0 | 27.4 |
| Lithuania | 6.655 | 4.645 | 11.300 | 11.300 | 35.415 | 46.7 | 18.7 | 76.5 |
| Luxemburg | 4.750 | 4.650 | 9.400 | 9.400 | 38.144 | 47.5 | 19.3 | 81.1 |
| Macau SAR | 10.124 | 11.316 | 21.440 | 21.440 | 28.725 | 50.2 | 31.8 | 57.5 |
| Macedonia | 7.845 | 8.010 | 15.855 | 15.855 | 30.305 | 46.2 | 20.4 | 66.2 |
| Madagascar | 18.208 | 18.804 | 37.012 | 37.011 | 8.703 | 45.7 | 57.2 | 18.1 |
| Malawi | 20.943 | 19.937 | 40.880 | 40.880 | 6.662 | 47.5 | 64.7 | 13.0 |
| Malaysia | 14.748 | 14.776 | 29.524 | 29.523 | 17.094 | 46.6 | 42.7 | 36.3 |
| Mali | 19.462 | 19.298 | 38.760 | 38.760 | 7.567 | 46.3 | 60.8 | 15.4 |
| Malta | 5.547 | 5.443 | 10.990 | 10.990 | 34.967 | 46.0 | 18.1 | 76.9 |
| Martinique | 8.689 | 8.934 | 17.623 | 17.624 | 31.568 | 49.2 | 23.0 | 64.5 |
| Mauritania | 21.459 | 20.542 | 42.001 | 42.001 | 5.522 | 47.5 | 67.2 | 10.5 |
| Mexico | 15.280 | 15.324 | 30.604 | 30.604 | 17.135 | 47.7 | 47.3 | 35.5 |
| Monaco | n/a | n/a | n/a | n/a | 45.311 | n/a | n/a | 100 |
| Mongolia | 16.471 | 16.644 | 33.115 | 33.115 | 15.216 | 48.3 | 49.6 | 31.0 |
| Montenegro | 7.221 | 6.533 | 13.754 | 13.754 | 32.440 | 46.3 | 18.3 | 70.8 |
| Morocco | 15.189 | 15.478 | 30.667 | 30.667 | 16.696 | 47.4 | 44.3 | 34.8 |
| Namibia | 17.746 | 17.865 | 35.611 | 35.612 | 11.084 | 46.7 | 54.5 | 22.1 |
| Nepal | 17.047 | 17.955 | 35.002 | 35.002 | 11.305 | 46.3 | 53.0 | 23.7 |

| Country | | | | | | | |
|---|---|---|---|---|---|---|---|
| Netherlands | 4.605 | 4.586 | 9.191 | 9.193 | 38.457 | 47.7 | 19.2 | 81.6 |
| New Zealand | 6.686 | 6.971 | 13.657 | 13.657 | 33.347 | 47.0 | 18.5 | 71.5 |
| Nicaragua | 18.974 | 19.019 | 37.993 | 37.993 | 9.525 | 47.5 | 58.9 | 19.2 |
| Niger | 20.003 | 21.372 | 41.375 | 41.376 | 5.286 | 46.7 | 66.0 | 10.2 |
| Nigeria | 18.260 | 20.077 | 38.337 | 38.337 | 7.723 | 46.1 | 59.7 | 15.8 |
| Norway | 4.107 | 4.449 | 8.556 | 8.556 | 37.713 | 46.3 | 17.7 | 82.4 |
| Panama | 12.047 | 13.149 | 25.196 | 25.195 | 22.353 | 47.6 | 33.7 | 46.9 |
| Pakistan | 16.132 | 17.055 | 33.187 | 33.187 | 12.933 | 46.1 | 49.3 | 27.4 |
| Paraguay | 14.551 | 15.521 | 30.072 | 30.072 | 15.845 | 45.9 | 43.2 | 30.9 |
| Peru | 14.36 | 15.039 | 29.399 | 29.399 | 17.685 | 47.1 | 41.8 | 37.2 |
| Philippines | 16.312 | 16.488 | 32.800 | 32.800 | 14.190 | 47.0 | 48.5 | 29.6 |
| Poland | 6.770 | 5.878 | 12.648 | 12.648 | 34.068 | 46.7 | 19.0 | 73.6 |
| Portugal | 4.584 | 4.161 | 8.745 | 8.745 | 37.867 | 46.6 | 17.9 | 82.1 |
| Puerto Rico | 8.065 | 7.723 | 15.788 | 15.789 | 30.973 | 46.8 | 44.1 | 66.7 |
| Qatar | 13.831 | 19.223 | 33.054 | 33.054 | 17.254 | 50.3 | 54.8 | 33.8 |
| Romania | 6.028 | 5.633 | 11.661 | 11.661 | 35.472 | 47.1 | 19.7 | 76.0 |
| Russia | 7.573 | 5.293 | 12.866 | 12.866 | 34.942 | 47.8 | 22.0 | 73.8 |
| Rwanda | 20.318 | 19.570 | 39.888 | 39.888 | 7.243 | 47.1 | 62.9 | 14.4 |
| San Marino | 2.796 | 3.847 | 6.643 | 6.643 | 41.742 | 48.4 | 19.9 | 87.3 |
| Saudi Arabia | 15.321 | 19.975 | 35.296 | 35.297 | 10.371 | 45.7 | 55.8 | 21.9 |
| Senegal | 18.381 | 19.533 | 37.914 | 37.915 | 8.056 | 46.0 | 58.9 | 16.6 |
| Serbia | 5.319 | 5.153 | 10.472 | 10.472 | 35.349 | 45.8 | 18.1 | 77.9 |
| Seychelles | 14.976 | 13.360 | 28.336 | 28.336 | 21.331 | 49.7 | 40.1 | 42.8 |
| Sierra Leone | 17.639 | 19.209 | 36.848 | 36.848 | 9.071 | 45.9 | 56.7 | 18.8 |
| Singapore | 9.583 | 10.629 | 20.212 | 20.211 | 31.880 | 52.1 | 32.4 | 61.5 |
| Slovakia | 7.003 | 6.092 | 13.095 | 13.095 | 33.656 | 46.8 | 18.7 | 72.7 |
| Slovenia | 5.822 | 4.509 | 10.331 | 10.331 | 37.408 | 47.7 | 19.5 | 79.1 |
| Somalia | 19.579 | 19.475 | 39.054 | 39.054 | 7.446 | 46.5 | 61.2 | 15.0 |
| South Africa | 14.869 | 13.895 | 28.764 | 28.764 | 18.703 | 47.5 | 40.7 | 39.1 |
| Spain | 4.238 | 4.087 | 8.325 | 8.326 | 39.902 | 48.2 | 19.5 | 83.6 |
| Sri Lanka | 10.712 | 11.600 | 22.312 | 22.312 | 25.204 | 47.5 | 28.7 | 53.2 |
| Sudan | 19.660 | 21.788 | 41.448 | 41.449 | 6.932 | 48.4 | 65.9 | 13.4 |
| Swaziland | 21.082 | 19.536 | 40.618 | 40.618 | 5.909 | 46.5 | 64.2 | 11.7 |
| Sweden | 2.782 | 3.156 | 5.938 | 5.938 | 39.741 | 45.7 | 17.4 | 88.1 |
| Switzerland | 3.699 | 3.664 | 7.363 | 7.363 | 40.084 | 47.5 | 19.2 | 85.4 |
| Syria | 18.382 | 19.419 | 37.801 | 37.802 | 9.696 | 47.5 | 58.5 | 19.5 |
| Taiwan | 7.923 | 10.088 | 18.011 | 18.011 | 30.382 | 48.4 | 24.3 | 63.2 |
| Tajikistan | 17.025 | 17.019 | 34.044 | 34.044 | 12.472 | 46.5 | 51.0 | 26.1 |
| Tanzania | 19.494 | 19.710 | 39.204 | 39.204 | 7.479 | 46.7 | 61.4 | 15.0 |
| Thailand | 10.727 | 11.165 | 21.892 | 21.892 | 26.824 | 48.7 | 29.8 | 55.2 |
| Togo | 20.972 | 20.122 | 41.094 | 41.094 | 5.227 | 46.3 | 65.3 | 10.2 |
| Tunisia | 12.624 | 13.817 | 26.441 | 26.442 | 21.545 | 48.0 | 36.2 | 44.7 |
| Turkey | 12.323 | 12.990 | 25.313 | 25.313 | 22.545 | 47.9 | 34.1 | 47.0 |
| Turkmenistan | 17.287 | 15.897 | 33.184 | 33.184 | 13.305 | 46.5 | 49.5 | 28.0 |
| UK | 3.681 | 4.048 | 7.729 | 7.729 | 38.399 | 46.1 | 17.9 | 84.2 |
| USA | 5.663 | 5.901 | 11.564 | 11.564 | 35.096 | 46.7 | 18.4 | 76.0 |
| Ukraine | 6.149 | 4.363 | 10.512 | 10.512 | 36.359 | 46.9 | 19.4 | 78.4 |
| UAE | 15.403 | 20.627 | 36.030 | 36.031 | 18.597 | 54.6 | 60.4 | 33.7 |
| Uganda | 22.226 | 23.084 | 45.310 | 45.310 | n/a | n/a | 74.1 | 0.00 |
| Uruguay | 7.142 | 6.716 | 13.858 | 13.859 | 31.990 | 45.9 | 17.4 | 70.4 |
| Uzbekistan | 16.231 | 15.448 | 31.679 | 31.679 | 14.892 | 46.6 | 46.6 | 31.4 |
| Venezuela | 13.944 | 14.378 | 28.322 | 28.322 | 19.001 | 47.3 | 39.9 | 39.9 |
| Vietnam | 15.087 | 13.916 | 29.003 | 29.003 | 18.520 | 47.5 | 41.4 | 38.7 |
| Yemen | 19.197 | 20.093 | 39.290 | 39.290 | 7.721 | 47.0 | 62.2 | 15.5 |
| Zambia | 22.572 | 21.859 | 44.431 | 44.431 | 5.137 | 49.6 | 71.8 | 9.3 |
| Zimbabwe | 18.239 | 19.082 | 37.321 | 37.321 | 10.872 | 48.2 | 57.6 | 21.8 |

n/a, data not available; 1, $K_{MT}$(m) is a $K_{QT}$ coefficient computed for the male populations, query object Monaco; 2, $K_{MT}$(f) is a $K_{QT}$ coefficient computed for the female populations, query object Monaco; 3, the sum of $K_{MT}$(m) and $K_{MT}$(f); 4, a $K_{MT}$ coefficient computed for the entire population pyramid, query object Monaco; 5, a $K_{UT}$ coefficient computed for the entire population pyramid, query object Uganda; 6, the sum of $K_{MT}$ and $K_{UT}$; 7, the sum of $K_{MT}$ and $K_{AT}$ where $K_{AT}$ is the coefficient for the entire population pyramid, query object Argentina; 8, $MU$ coefficient, see Eq. (3).

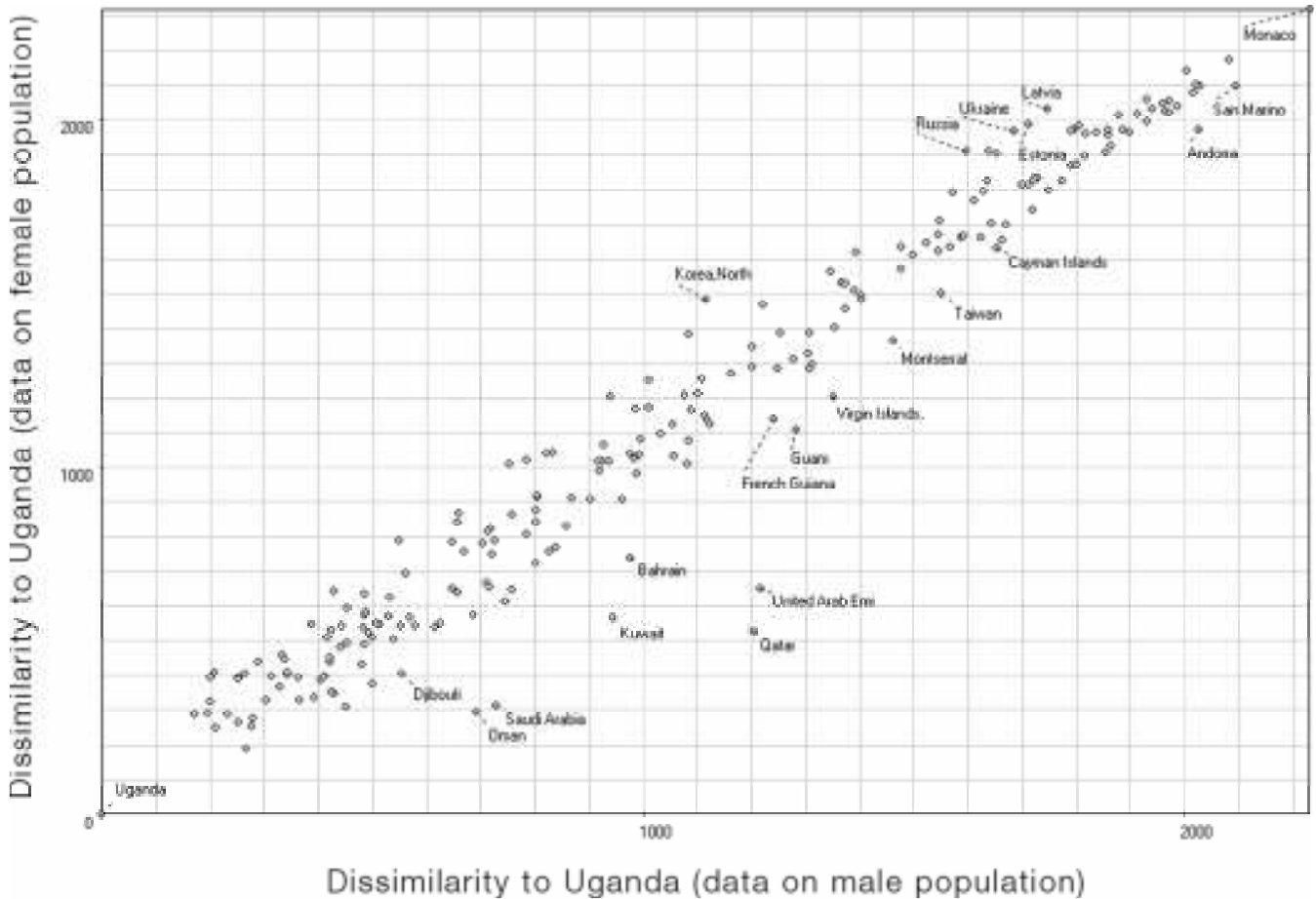

Fig. 2. Correlation between dissimilarity coefficients $D_{UT}$(f) и $D_{UT}$(m) for population pyramids of 220 countries (data source: U.S. Census Bureau, International Data Base, IDB Summary Demographic Data, 2000. http://www.census.gov/ipc/www/idbsum.html). Δ = 0.01.

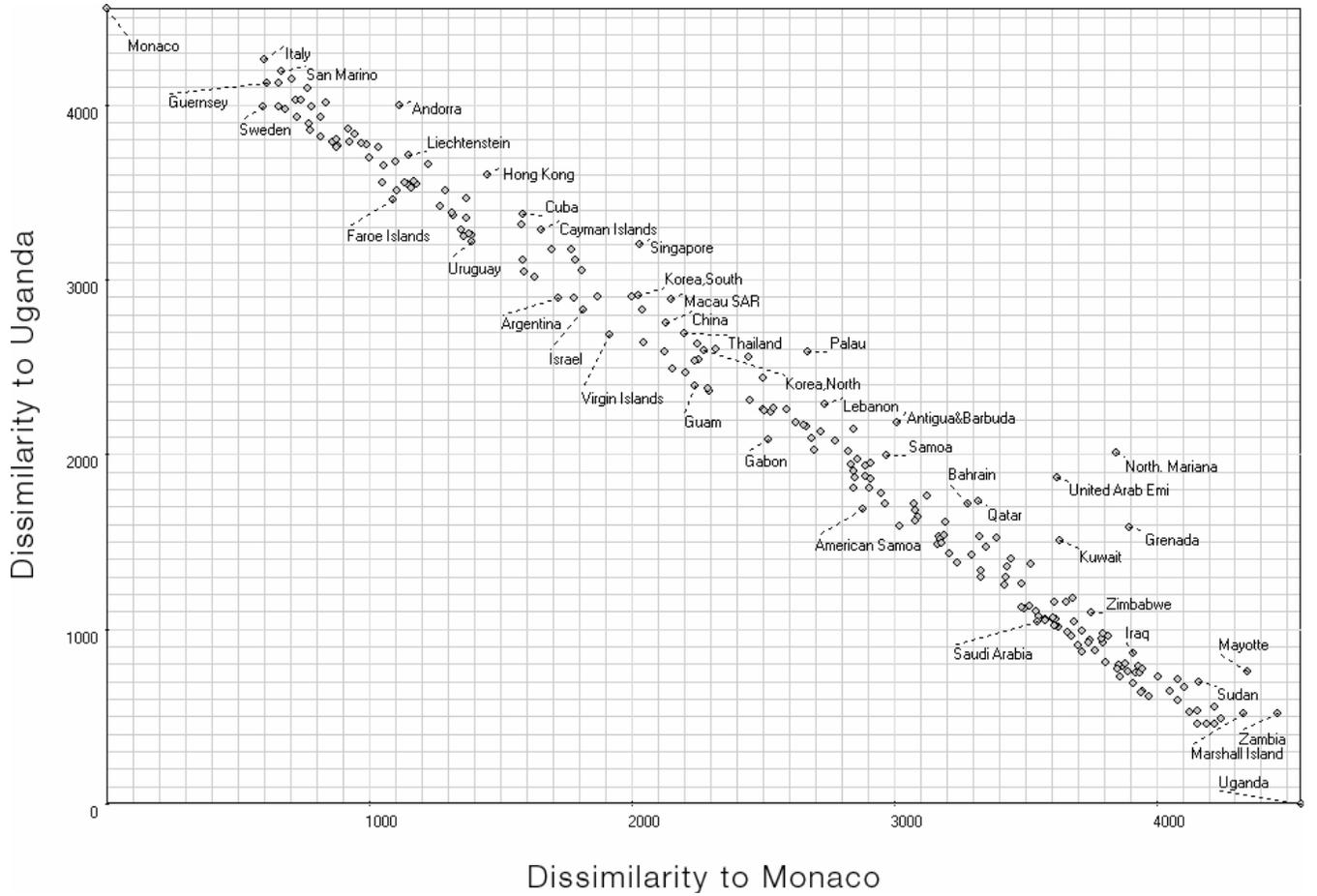

FIG. 3. Correlation between dissimilarities of population pyramids to Uganda's pyramid and to Monaco's pyramid. Δ = 0.01.

dependence between the values of dissimilarity coefficients computed for 220 countries upon the use query objects Uganda and Monaco, which demonstrates the consistency of the sum of $K_{UT}$ and $K_{MT}$.

The additivity of the $K_{QT}$ coefficients allows the computation of an index that shows a percent value of similarity between population pyramids. Such an index may appear very informative in finding correlations between population pyramids and characteristics of populations. Index $MU$, showing percent of similarity of a target country population pyramid to the Monaco population pyramid is computed by the formula:

$$MU_T = \frac{100 \times K_{UT}}{K_{UT} + K_{MT}} \qquad (3)$$

The $MU_T$ values are shown in Table 2, column 8. Thus, the $MU$ index reflects the balance between two polar "viewpoints" on the global entirety of the population pyramids. The two polar viewpoints are provided by the computer ego created based on population pyramids of Uganda and Monaco. Notably, in case of relatively well-doing countries with high GDP per capita, developed democracy, reasonably good healthcare, low infant mortality, and other characteristics of higher standard of life, the $MU$ index exceeds 70, whereas the same index for the countries with deep problems in the above-said areas is less than 25. Out of 220 countries, the former constitute 24.5%, and 85% of them are European countries. The countries with $MU$ index lower than 25 make 30% of the whole list of 220 countries, and 76% of such countries are located on the African continent. These findings are especially interesting because the $MU$ indexes were determined based on absolutely objective data, such as population composition by age and sex groups.

## 3.4. Use of $K_{QT}$ coefficients in modeling

As the age and gender group distribution in a population pyramid is clearly not the cause but an effect of the standard of life in a respective country, the understanding of the mechanism of formation of population pyramids is extremely important. The additivity of $K_{QT}$ coefficients allows construction of various models that can help understand the nature of the above-noted differences in population pyramids of various countries. A comparison of the population pyramids shown in Fig. 1, for instance, of Monaco and Sweden, on the one side, and Uganda and Angola, on the other side, leads to understanding of the major difference between these two pairs of countries that have the most antipodal $K_{QT}$ indexes. The former have uniform shapes, i.e. the distribution of different age cohorts in the total population is close to even. In the second pair of countries, the shares of older age cohorts in the total populations exponentially decline. The most natural explanation for this regularity seems to be that in developed countries the mortality rate does not greatly depend on the age, whereas in economically and otherwise challenged countries the mortality rate among older groups of population is higher. One way or the other, it is easy to construct population pyramid models reflecting a certain viewpoint and verify whether the model-based data agree with the experimental data.

In this study, we constructed a hypothetical uniform (UN) population pyramid in which each age cohort was represented equally and had a share of 1/34 of the total population. We also constructed two exponential type pyramids, *E20* and *E30*, which were computed in a same way but separately for male and female populations of each of the pyramids and so that the share of each of the successive age cohorts was lower than a previous cohort by 20 and 30%, respectively. For instance, the share of the 0-4 age cohort in the 30%-model is 30.07%, whereas the share of the 5-9 age cohort was 21.05%. Then we computed dissimilarity coefficients of each of 220 population pyramids to the uniform and exponential models, which showed a very distinct linear correlation. The exponential model with 30% decrease in the age cohort shares appeared to show less spread data than the 20%-model. This is clearly seen on Fig. 4 that shows the correlations between the countries' *MU* indexes and their dissimilarities to models *E20* and *E30*, on the example of 162 countries.

The computer ego modeling used in this study showed that all the existing diversity of population pyramids can be represented in the form of an additive combination of two patterns – the uniform and exponential ones. The average value of the sum of dissimilarities of 220 population pyramids to UN and *E30* individually is 64.50 with an average deviation of 1.47. The values of $D_{UN,T}$ and $D_{E,T}$ vary by 6.4 and 3 times, respectively. These results provide for computation of a share of the uniform component, $P_{UN}$, of any population pyramid according to the formula:

$$P_{UN,T} = \frac{100 \cdot (1 - D_{UN,T})}{D_{UN,T} + D_{E30,T}} \tag{4}$$

$P_{UN,T}$ linearly correlates with *MU*. In the population pyramid of Monaco, the share of the uniform pattern is 96.3%, while for Uganda it is 34.6%.

## 3.5. Demographic correlations

An intelligent method for pattern recognition applied to population pyramids should provide a capability to discover correlations between the regularities in distribution of sex and age groups and other demographic characteristics of a population. In this respect, the *MU* index has an advantage over the traditional set of 34 parameters. First of all, it represents a holistic characteristic of the population of a given country; secondly, it represents a holistic characteristic of the entire world population; and thirdly, it measures progressive tendencies in the development of a country's population. A higher *MU* index reflects a higher standard of life in a given country. This is clearly seen from the examples of a few correlations demonstrated below.

We will start with a population characteristic that may seem to have no relation to the shape of a population pyramid – a national IQ score. Not long ago, it was convincingly demonstrated by Richard Lynn and Tatu Vanhanen [8] that the assumption of equality of average intelligence in different nations was erroneous. On the examples of 185 countries, the authors have shown that average national IQ scores significantly vary, and the average national IQ of the world is 90. As demonstrated in Fig. 5, we have established that countries with *MU* index above 50 have national IQ scores above 90; and, conversely, *MU* index below 50 correlates with IQ below 90. As is seen, along with the generally steady correlation (shaded area of the graph), there are two groups of notable deviation: 1) countries with IQ higher than the general correlation pattern; and 2) countries with lower than correlation-based values of IQ. The first group includes the countries of East and Southeast Asia, and the second group mainly consists of island states. While the interpretation of this finding is certainly beyond the scope of this publication on a new method for pattern recognition, the fact itself is important both in the context of demonstration of the method and as a part of the IQ correlations issue. It is possible that the correlation between the average national IQ score and the *MU* index is indirect and is due to the fact that a higher standard of life is conducive to higher level of education in the country.

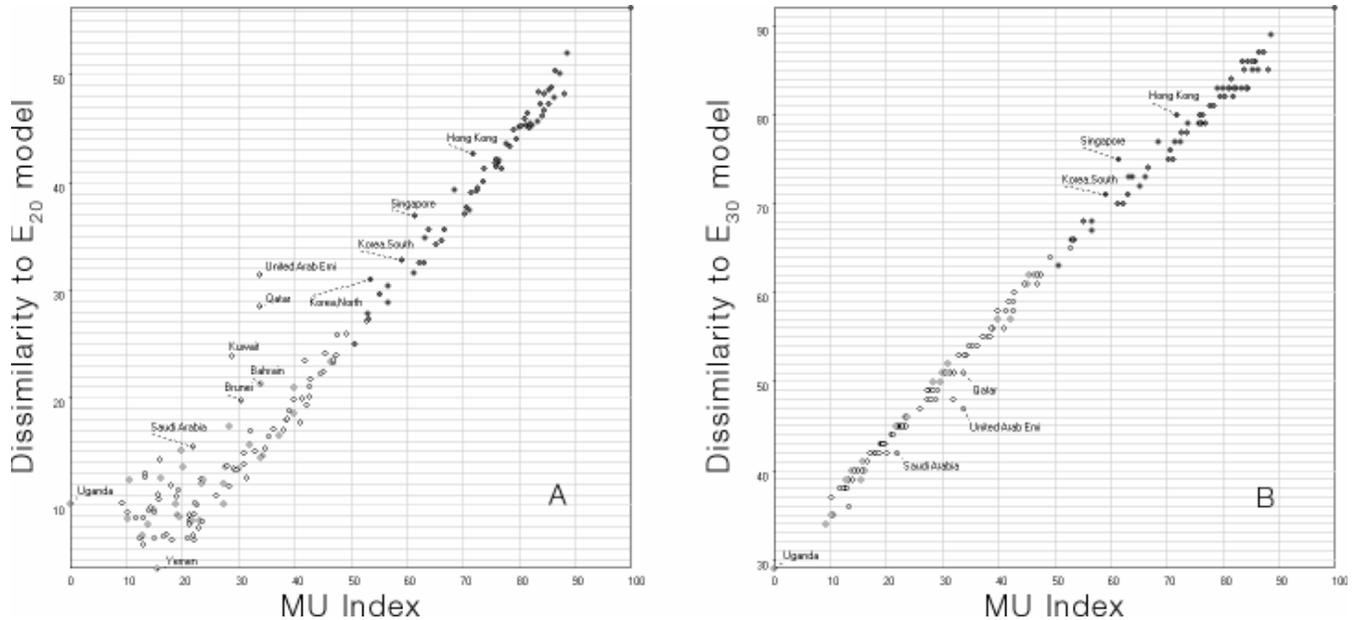

FIG. 4. Correlations between dissimilarities of population pyramids to the exponential model of age group distribution and *MU* index (for 162 countries). The *E20* population pyramid model (Fig. 4A) is based on 20% decrease in populations of each 5-year older age group, and the *E30* model (Fig. 4B), on 30%.

Unlike a national IQ score, a GDP per capita directly reflects the welfare of a nation and should be proportional to the *MU* index. Fig. 6 shows a distinct correlation between logarithms of GDP per capita and MU indexes. Similarly to the above demonstrated, here, too, along with the general correlation between the two indexes, there are two groups of exceptions; however, of a different nature. The data points located above the shaded area of the steady general correlation between *MU* index and GDP per capita that demonstrates the exponential dependency between the GDP and *MU* indexes, is characteristic of the countries with extraordinarily vast sources of income: for instance, major exporters of petroleum; Bahamas, with the economy based on income from well-established tourism and financial services; Botswana whose economy is one of the most dynamic in Africa due to extensive nature preserves and stable social progress; Singapore, one of the world's most prosperous countries; etc.

Beneath the shaded area are the countries whose economies have been in one way or another affected by the consequences of the Communist ideology and, as such, are different from the countries with established market economies.

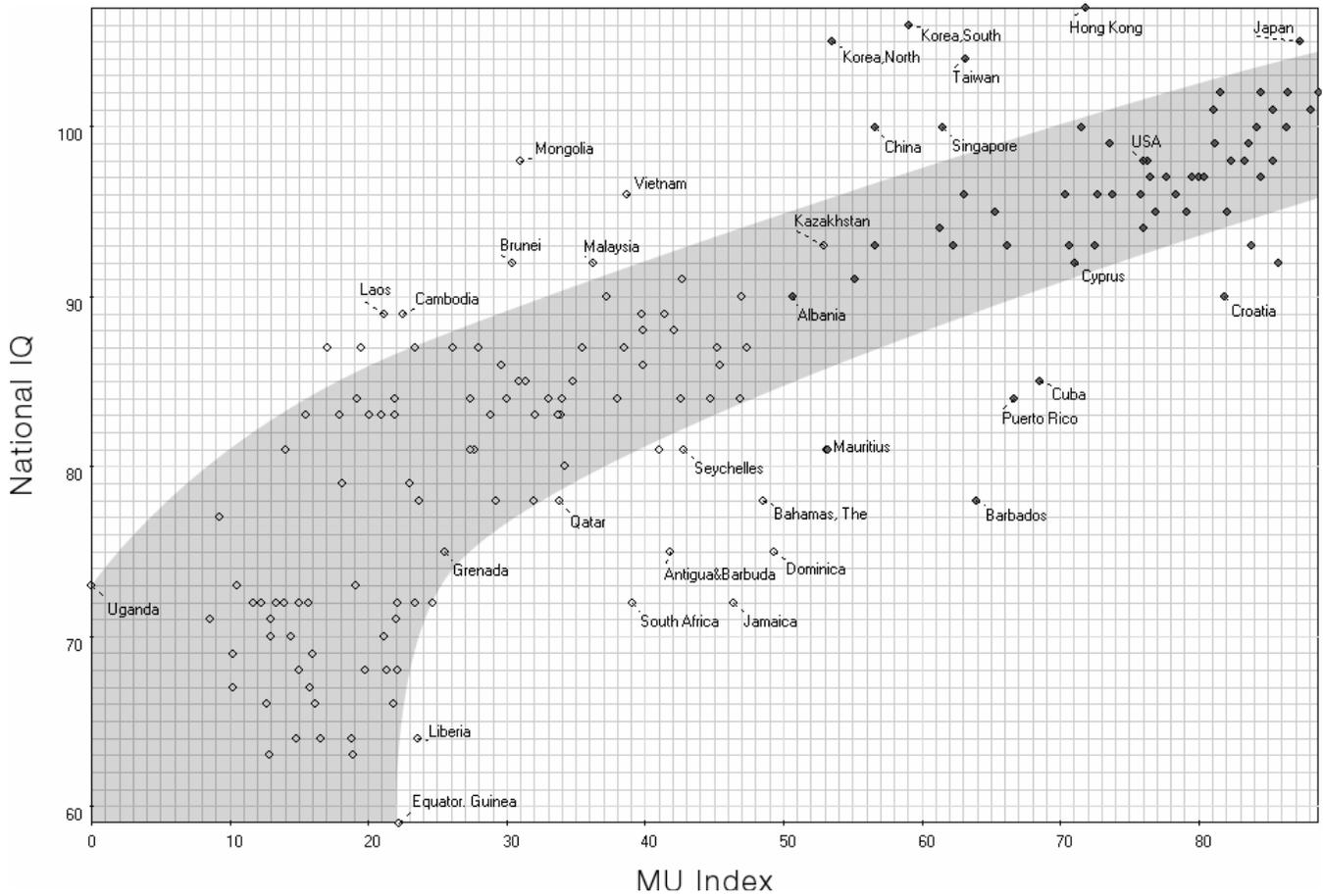

FIG 5. Correlation between average national IQ scores [8] and *MU* indexes. (data source: U.S. Census Bureau, International Data Base, IDB Summary Demographic Data, 2000. http://www.census.gov/ipc/www/idbsum.html).

Three factors that directly influence the formation of population pyramids are: birth and death rates and life expectancy. Migration rate has a less effect as it is government regulated. The relation between *MU* and death rate is quite complex as is seen on Fig. 7. Fig. 7 shows two curves – a steep curve at low *MU* index values and a smooth one at *MU* index values of 30 and above – that gradually join and become one curve. The area where the two curves join corresponds to approximately equal shares of the uniform and exponential components of the population pyramids. The dynamics of the first half of the curve is understandable and is explained by the fact that as a nation's welfare grows, correlating with higher *MU* values, the death rate goes down. As far as the subsequent gradual increase in the death rate correlating with higher *MU*, it can be attributed to a combination of many indirect factors. Detailed comparative studies of demographic situations, including analyses of factors contributing to the death rate are available elsewhere and cover various countries, for instance, post-communist Russia and Ukraine [10, 11] that have the highest deviations from the "normal" '*MU* - death rate' correlation (see Fig. 7). However, a detailed analysis of a particular country's population would hardly be relevant in these findings that provide a "bird's eye view" of dynamics of the death rate for the entire human population of the world. To explain the correlations found by us, it should be more productive to look in the direction of generalization of the peculiarities of nation's populations which manifest themselves in deviations from the main tendency. In this particular example, such peculiarities are clearly visible: all of the countries that display a significant tendency towards a higher death rate in the situation of a smooth correlation between *MU* and death rate shown in Fig. 7 are the former USSR republics and the east European countries of the former Soviet bloc. Thus, it can be concluded that the above peculiarities closely correlate with historical- political factors and that the analysis of those peculiarities should take into consideration the said factors.

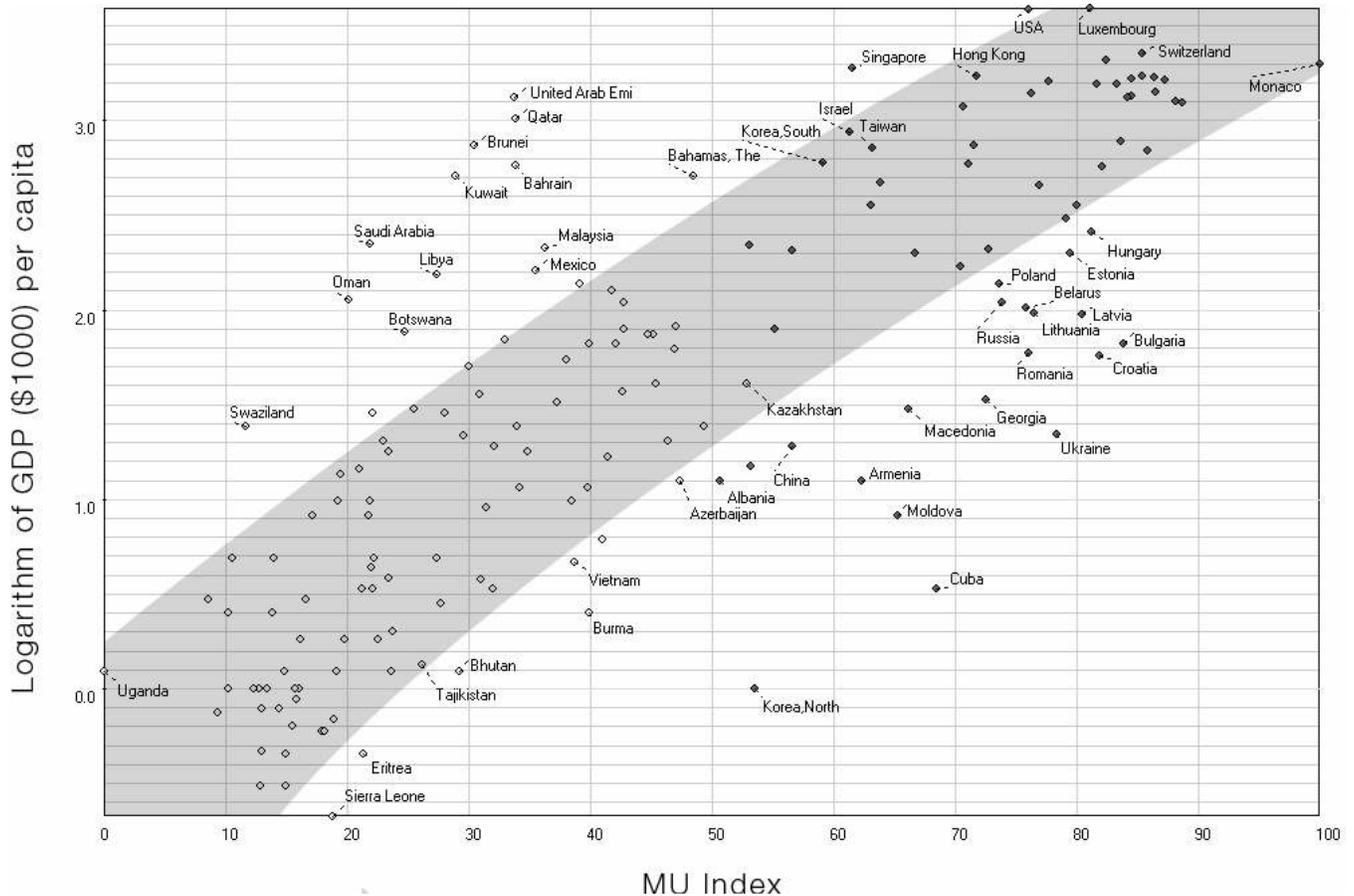

FIG 6. Correlation between the MU index and logarithm of gross domestic product per capita, in $1000 [9].

At first glance, a correlation between fertility (Fig. 8), hence birth rate (Fig. 9), and the *MU* index is simple and clear-cut: birth rate exponentially declines as welfare grows. However, a more complex and fine structure of the relationship between birth rate and the growth of the *MU* index is revealed upon analysis of dynamics of an index that is the reverse of the birth rate, i.e. population per birth (PPB) which reflects the number of a countries population per one newly born child. The relationship between PPB and *MU* is shown in Fig. 10.

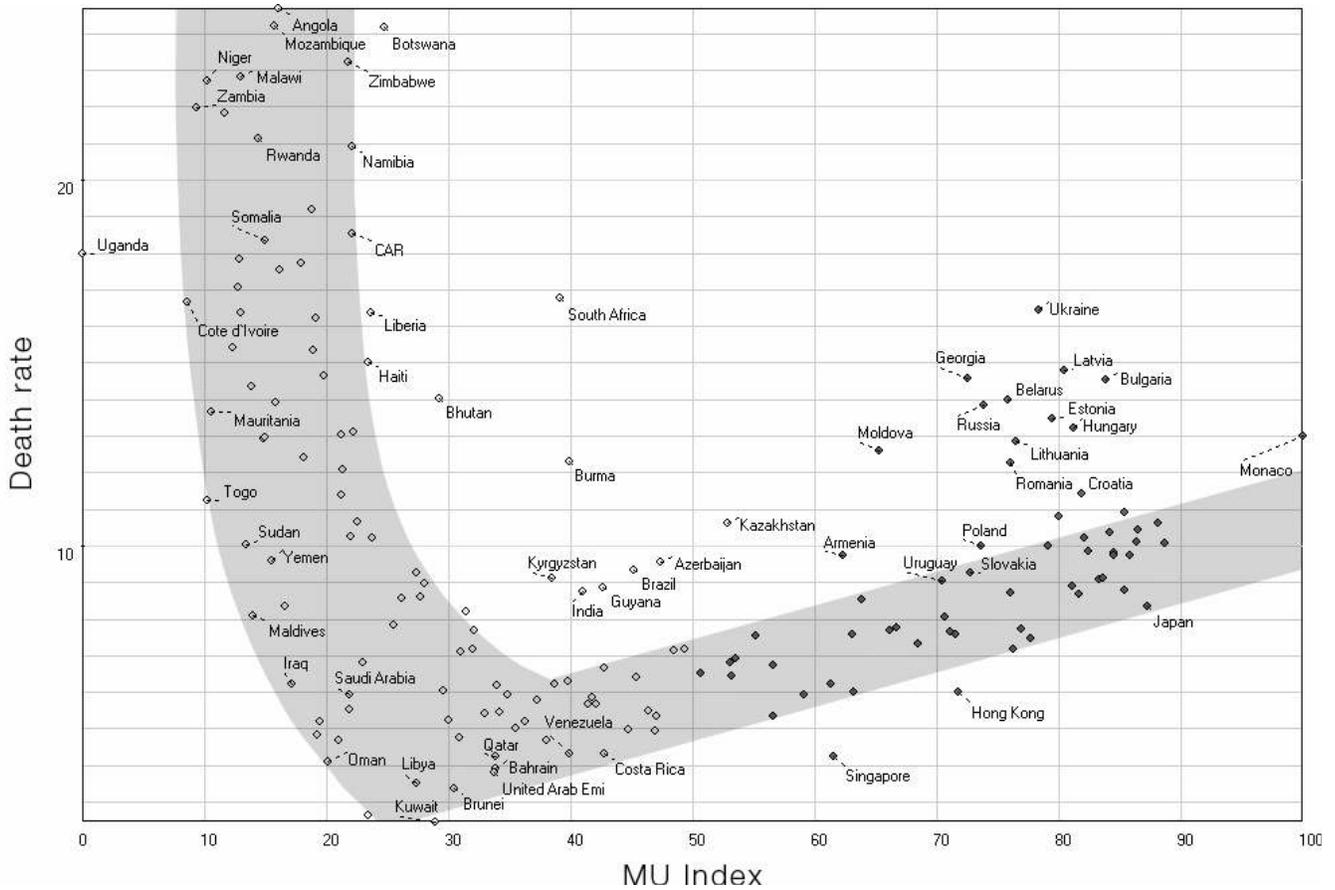

FIG. 7. Correlation between death rate (data source: U.S. Census Bureau, International Data Base, IDB Summary Demographic Data, 2000.http://www.census.gov/ipc/www/idbsum.html) and *MU* index.

As is seen, in most of cases of 162 population pyramids, the PPB growth practically linearly correlates with the growth of the *MU* index. There are three clearly visible deviations from linearity. One of them is observed at *MU* values slightly higher than 30 and can be explained by increased numbers of migrant laborers involved in oil-recovery industry, who, a rule, are temporary residents. The other two deviations towards higher PPB values are observed for the former USSR republics and the former Soviet bloc countries (at *MU* values of 70-80), as well as European countries with high standard of life and social security (*MU* values of about 85). These deviations from the linear dependency are very distinct and should be of interest to professional demographers. Unlike the birth rate, PPB carries a highly informative demographic characteristic: how many people per birth are involved in creation of the environment to which a newly born child arrives. This environment includes not only parents and families of newly born population but also the labor force involved in a child's healthcare, education, and relevant infrastructure and industries (infant food, clothing, toys, educational products, social programs for family and child support, etc.).

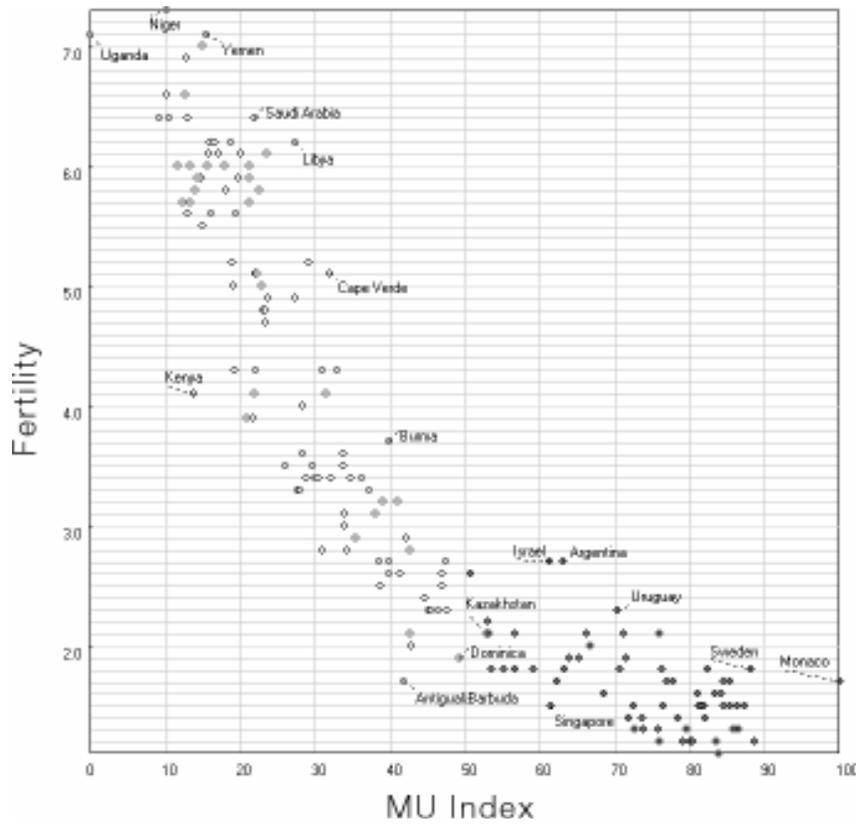

FIG. 8. Correlation between fertility and *MU* index.

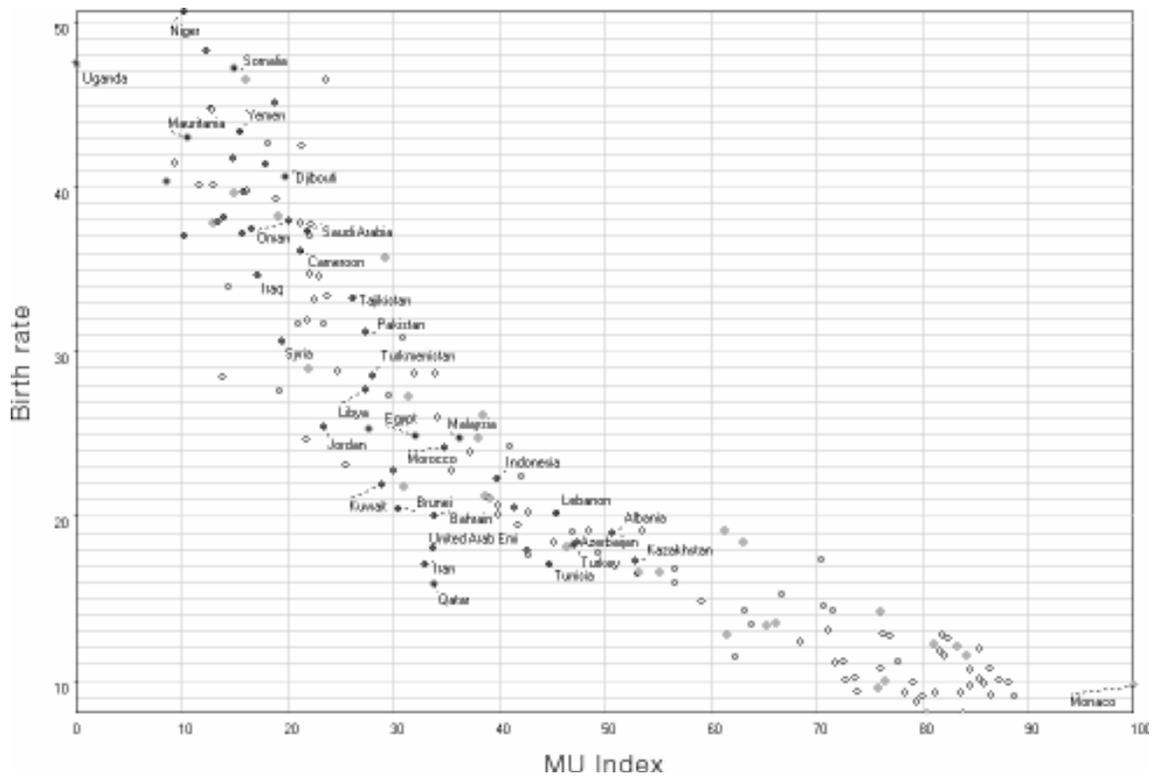

FIG. 9. Correlation between birth rate (births per 1000 persons) and *MU* index (for 162 countries).

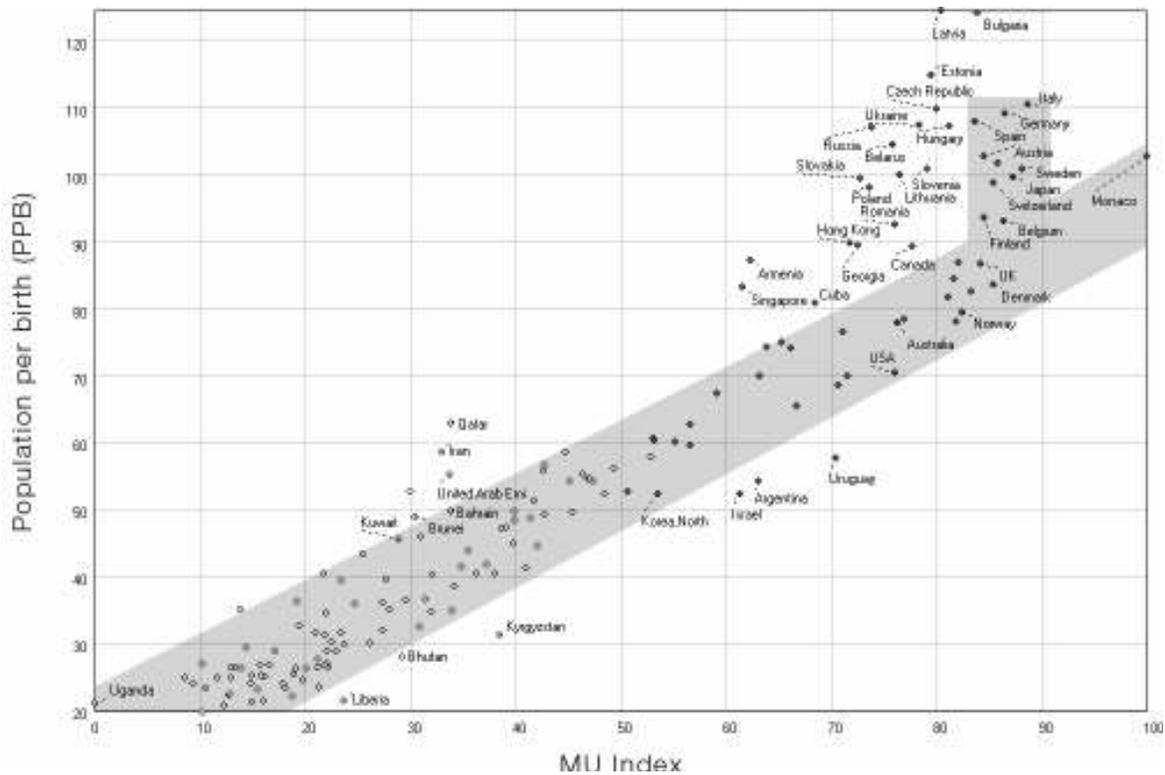

FIG. 10. Correlation between population per birth (PPB) and *MU* index for 162 countries.

## 4. Conclusions

By using the example of analysis of 220 population pyramids, we have provided a detailed description of the HGV2C method that represents a modification of the previously described HGV method [4]. The HGV2C modification is most simple for implementation and is based on the use of the computer ego which is created based on an individual object. It should be emphasized that this study was not aimed at demographic research and that the population pyramids were used as an object of analysis with the purpose of demonstration of the capabilities provided by the HGV2C method. As it is clear from the foregoing, the HGV2C method can be effectively used for recognition of any kinds of patterns. More complex examples of application of the HGV method were provided in [4].

The involvement of computer ego implemented through the use of a hypothesis-parameter and infothyristor contributes the intelligence factor to the data processing as it goes beyond the zero-reader approach when a method is intended only to detect the presence or absence of target objects. The HGV2C method provides the evaluation of an entire dataset from a position of an individual observer. One of the advantages of such an individualized perception of the nature of patterns under analysis is the capability to consolidate a totality of data through comparison of two opposite points of view on the nature of phenomena under analysis, as, for instance, was done in the above example by using a computer ego based on two fundamentally different population pyramids of Uganda and Monaco. The capability to easily model any query objects for construction of computer ego provides the means to join all the population pyramids into a certain harmonious system that can be described by one common criterion, instead of 34 individual parameters. In the above-demonstrated example the common criterion is the *MU* index that reflects the shares of the uniform and exponential components in each of the population pyramids.

This provides further new opportunities for discoveries of various demographic correlations and new approaches to investigation of various factors involved in formation of age and sex components of population pyramids.

An important peculiarity of the HGV2C method is the additivity of the response. A total value of dissimilarity (similarity) between a query and target objects equals the sum of contributions of individual parameters, i.e. increments of dissimilarities according to each individual parameter. This provides a capability to transform an array of data on a complex set of objects with similar genesis processes into a certain indivisible whole, and, based on that whole, to analyze the contribution and functional value of each of its elements.

The infothyristor that is employed by the HGV2C method as a special element of information processing is based on the previously described phenomenon of iterative averaging [1]. However, unlike the method based solely on iterative averaging, it allows comparative analysis of open systems when the composition of a database under analysis can be changed in the course of data processing, thus allowing studies into dynamics of behavior of complex systems described by an unlimited number of parameters.

**Acknowledgments**

**References**
[1] L. Andreev From a set of parts to an indivisible whole. Part I. Operations in a closed mode. arXiv:0803.0034 [cs.OH].
[2] L. Andreev. Unsupervised automated hierarchical data clustering based on simulation of a similarity matrix evolution. U.S. Patent 6,640,227 (2003).
[3] L. Andreev. High-dimensional data clustering with the use of hybrid similarity matrices. U.S. Patent 7,003,509 (2006).
[4] L. Andreev and M. Andreev. Method and computer-based system for non-probabilistic hypothesis generation and verification. U.S. Patent 7,062,508 (2006).
[5] O. Duda, P. E. Hart, D. Stork. Pattern Classification (2nd Edition), Wiley 2002
[6] C.Bishop. Pattern Recognition and Machine Learning. Springer, 2006.
[7] D. Maltoni, D. Maio, A. K. Jain, S. Prabhakar. Handbook of Fingerprint Recognition. Springer, New York, 2003.
[8] R. Lynn and T. Vanhanen (2002). IQ and the Wealth of Nations. Westport, CT: Praeger.
[9] CIA – The World Factbook, 2000. http://www.cia.gov/cia/publications/factbook/ census.gov/ipc/www/idbsum.html
[10] E. Brainerd and D. M. Cutler. Autopsy on an Empire: Understanding Mortality in Russia and the Former Soviet Union. *Journal of Economic Perspectives*, American Economic Association, vol. 19(1), pages 107-130, 2005.
[11] G. A. McDermott. Embedded Politics: Industrial Networks and Institutional Change in Post-communism. Ann Arbor: University of Michigan Press, 2002.